# Nonequilibrium transport and the fluctuation theorem in the thermodynamic behaviors of nonlinear photonic systems

Yang Liu[1], Jincheng Lu[2], Zhongfei Xiong[3], Fan O. Wu[4], Demetrios Christodoulides[5], Yuntian Chen[3], and Jian-Hua Jiang[6, 7, 1, †]

[1]*School of Physical Science and Technology & Collaborative Innovation Center of Suzhou Nano Science and Technology, Soochow University, 1 Shizi Street, Suzhou 215006, China*

[2]*Jiangsu Key Laboratory of Micro and Nano Heat Fluid Flow Technology and Energy Application, School of Physical Science and Technology, Suzhou University of Science and Technology, Suzhou 215009, China*

[3]*School of Optical and Electronic Information, Huazhong University of Science and Technology, Wuhan 430074, China*

[4]*CREOL, College of Optics and Photonics, University of Central Florida, Orlando, Florida 32816-2700, USA*

[5]*Ming Hsieh Department of Electrical and Computer Engineering, University of Southern California, Los Angeles, California 90089, USA*

[6]*School of Biomedical Engineering, Suzhou Institute for Advanced Research, University of Science and Technology of China, Suzhou 215123, China*

[7]*School of Physics, University of Science and Technology of China, Hefei 230026, China*

[†]*Corresponding author: jhjiang3@ustc.edu.cn*

**Abstract**

Nonlinear multimode optical systems have attracted substantial attention due to their rich physical properties. Complex interplay between the nonlinear effects and mode couplings makes it difficult to understand the collective dynamics of photons. Recent studies show that such collective phenomena can be effectively described by a Rayleigh–Jeans thermodynamics theory which is a powerful tool for the study of nonlinear multimode photonic systems. These systems, in turn, offer a compelling platform for investigating fundamental issues in statistical physics, attributed to their tunability and the ability to access negative temperature regimes. However, to date, a theory for the nonequilibrium transport and fluctuations is yet to be established. Here, we employ the full counting statistics theory to study the nonequilibrium transport of particle and energy in nonlinear multimode photonic systems in both positive and negative temperature regimes. Furthermore, we discover that in situations involving two reservoirs of opposite temperatures and chemical potentials, an intriguing phenomenon known as the loop current effect can arise, wherein the current in the positive energy sector runs counter to that in the negative energy sector. In addition, we numerically confirm that the fluctuation theorem remains applicable in optical thermodynamics systems across all regimes, from positive temperature to negative ones. Our findings closely align with numerical simulations based on first-principles nonlinear wave equations. Our work seeks to deepen the understanding of irreversible non-equilibrium processes and statistical fluctuations in nonlinear many-body photonic systems which will enhance our grasp of collective phenomena of photons and foster a fruitful intersection between optics and statistical physics.

## Introduction

The emergence of thermodynamic behaviors in many-body systems is remarkable in the sense that the huge gap between deterministic dynamics and statistical thermodynamics is still not filled after centuries study. In fact, such a topic recently attracts research interest again, thanks to the appearance of several systems where the many-body dynamics can be computed numerically [1-6]. One of these systems is the interacting bosonic system. When the interaction is not too strong, the system tends to reach equilibrium states with emergent thermodynamic behaviors that can be tuned by microscopic properties such as the dispersion as well as macroscopic properties such as the total energy and particle number. Such tunable emergent thermodynamic behaviors provide promising opportunities for the study of the fundamental issues in statistical thermodynamics.

One prototype is the multi-mode optical systems (MMOSs) in various lattices which offer an extremely vigorous platform to study the collective phenomena of photons. Over the past decades, MMOSs have been investigated extensively in their linear optical properties such as transmission in the presence of disorders [7-9], topological phenomena [10-15] ranging from Floquet quantum Hall effects [16-21] to Dirac and Weyl physics [22-26], and non-Hermitian effects [27-29]. Meanwhile, nonlinear effects in MMOSs such as parametric instabilities [30-33], spatiotemporal mode-locking [34, 35], and multi-mode solitons [36] have also been explored, revealing intriguing phenomena that are unavailable in single-mode or few modes optical systems. These linear and nonlinear effects are also promising for various applications in photonics.

Indeed, there are interesting phenomena in nonlinear multi-mode optical systems (NLMMOSs) that can be related to thermodynamics. It was observed recently that in nonlinear multi-mode optical fibers, driven by the Kerr nonlinear effect, the optical power resides in the various optical modes at the input can automatically evolve to an output state that favors only a few lower-order modes. This effect, known as the beam self-cleaning mechanism [32, 37, 38], was recently revealed as a manifestation of optical thermalization effect which is universal in NLMMOSs [39, 40]. In other words, thermalization processes in multimode nonlinear photonic systems lead to equilibrium states with Rayleigh-Jeans distributions that resemble beam self-cleaning effects at the output. It is worth noting that the emergent Rayleigh-Jeans distribution contains an optical temperature and a chemical potential that can be largely tuned by the system's total energy and particle number as well as the mode couplings [41-48]. This tunability even gives rise to the emergence of negative temperatures that have been observed in recent experiments [49]. Therefore, NLMMOSs offer an excellent platform to study the fundamental statistical physics as well as a desirable material system for various optical applications.

Although the thermodynamic theory for photons in NLMMOSs has been established [41, 46], the theory and experiments for nonequilibrium transport remain largely unexplored. Transport and fluctuations are fundamental phenomena when multiple NLMMOSs exchange energy and particle with each other. In this work, we promote the theory for nonequilibrium transport and fluctuations from two complementary approaches. We use the full counting statistics (FCS) theory based on quantum master equation to study the particle transport and fluctuations between two NLMMOSs. Meanwhile, we employ the nonlinear Schrödinger equation to numerically simulate the photon dynamics in the combined system. Such numerical simulation can provide the transport current and its fluctuation dynamics in the system which can be used to extract the average current and the current statistics. These properties are then compared with the theoretical results from the FCS theory. We find reasonable agreement between the two approaches, although they have different starting points (the simulated systems haven't reached thermodynamic limit yet, due to their finite sizes). We then use these two approaches to study transport and fluctuations in various regimes. These complementary approaches can also be used to verify the fluctuation theorem in the negative temperature regime. Remarkably, we theoretically find that the fluctuation theorem holds even for such negative temperature regime. These results are further validated by the numerical simulations. In particular, the heat exchange fluctuation theorem [50, 51] is confirmed for both positive and negative temperatures by considering transport between two NLMMOSs coupled by a cross-phase modulation (XPM) mechanism. Our work paves the way for the development of a comprehensive nonequilibrium statistical physics theory for nonlinear photonic systems.

**Thermodynamic theory for photons in NLMMOSs**

We would like to first introduce two examples of optical thermodynamics in NLMMOSs (Fig. 1). The first example is arrays of cavities or waveguides [52-57]. As shown in Fig. 1(b), a two-dimensional (2D) array of coupled ring cavities can serve as highly tunable NLMMOSs. Here, both the lattice structure and the mode couplings can be tuned by the geometry of the system, while the nonlinearity can be tuned by the materials that form the system. The second example is nonlinear multi-mode optical fibers which are characterized by refractive index profiles $n(x, y)$ that guide photons' propagation along the $z$ direction (such a direction is often regarded as the time evolution axis in Schrödinger's equation). It has been revealed that chaotic nonlinear dynamics in this type of systems can lead to thermalization effects (e.g., Rayleigh–Jeans distributions and mode self-cleaning) [58-60].

According to Ref. [41], the thermodynamic behaviors of the photonic system in NLMMOSs can be described by introducing a number of thermodynamic quantities. First, the optical power $P$ is defined as the sum of the mode occupancy coefficient ($a_i = \langle \psi_i | \Psi \rangle$) as given by $P = \sum_i^M |a_i|^2$. We use the analog of quantum states and the Dirac notation to formulate the theory as the system is essentially described by a nonlinear Schrödinger equation. Here, $i$ denotes the mode index, $\psi_i$ is the $i$-th eigenstate of the photonic system without the nonlinear effect, $M$ is the total number of such eigenstates, and $|\Psi\rangle$ stands for the photonic state of the entire system. Therefore, $P$ can be regarded as related to the total photon numbers which is preserved in the absence of gain and loss if we consider Kerr nonlinearity. Another invariant is the total energy of the photonic system as represented by the Hamiltonian which can be divided into the linear and nonlinear parts, $H = H_L + H_{NL}$. The weight of the nonlinear Hamiltonian $H_{NL}$ depends on the optical power and the strength of the nonlinear coefficients. Under the weak nonlinearity assumption, the system Hamiltonian is dominated by its linear part, $H \approx H_L$. Therefore, the internal energy $U$ can be defined as $U = \sum_i^M \varepsilon_i |a_i|^2$ where $\varepsilon_i$ is the eigen-energy of the $i$-th eigenstate. Here, it is worth remarking that throughout this paper, we adopt the notation specific to coupled optical cavity systems that evolve over time. In contrast, for multimode optical fibers, the definition of energy usually involves a minus sign since there the fundamental mode often has the largest quasi-energy eigenvalue. Apart from such difference, the framework of the theory is very much the same for coupled cavity systems and multimode optical fibers. For a detailed comparison of notations, please refer to [41].

The nonlinear effect here is crucial for the energy exchange among the different eigenmodes. Such energy exchange drives the system to ergodically go through all the microstates on the constant power and energy manifolds a fair manner, like in microcanonical thermodynamic ensembles. This aspect has been numerically verified for various lattice types, dimensions, and nonlinear effects and is a universal behavior in NLMMOSs when the nonlinearity is not too strong. Since photons are boson and the number of photons is much greater than the number of modes, maximization of entropy leads to a special case of the Bose-Einstein distribution—the Rayleigh-Jeans distribution. In this case, the Gibbs entropy $S$ of the system can be written as $S = M + \sum_{i=1}^{M} \ln|a_i|^2$ (see Appendix A for the derivation; In this work we set $k_B = 1$) [43]. At thermodynamic equilibrium, the mode occupancy coefficient at thermodynamic equilibrium state is found as [43]

$$|a_i|^2 = \frac{T}{\varepsilon_i - \mu}. \tag{1}$$

To directly visualize the above thermalization process, we calculate the evolution of the mode occupancy coefficient with different initial conditions. The model system is a uniform square lattice with $20 \times 20$ sites and the coupling $\kappa$ between the nearest sites is set as unit 1. The corresponding Hamiltonian operator takes the following form

$$\hat{H} = \sum_{<i,j>} \kappa \hat{c}_i^\dagger \hat{c}_j + \sum_i \chi^{(3)} \hat{c}_i^\dagger \hat{c}_i \hat{c}_i^\dagger \hat{c}_i, \tag{2}$$

where $<i,j>$ denotes the nearest-neighboring sites. That is, the linear part of the Hamiltonian is determined by the nearest-neighbor couplings, while the nonlinear part is governed by, e.g., the Kerr nonlinearity. The equation of motion is given by the nonlinear Schrödinger equation (see Appendix B for the procedure towards dimensionless quantities):

$$i\frac{da_i}{dt} + \sum_{<i,j>} \kappa a_j + \chi^{(3)}|a_i|^2 a_i = 0, \tag{3}$$

where $a_{i/j}$ is the time-dependent wavefunction at the site $i/j$ of the NLMMOS. Note that $\hbar$ is set as 1 in this work for simplicity. Based on this equation, one can simulate the thermalization process with various initial conditions. As shown in Figs. 1(c) and 1(d), the photonic system reaches the thermal equilibrium state after a short period of time, visualizing directly the thermalization process. During such processes, the system's entropy increases until it reaches its equilibrium state.

It is worth mentioning that in order to simulate the dynamics of a microcanonical ensemble, we numerically integrated the above nonlinear Schrödinger equation numerically for 8000 times with the same initial mode distribution $|a_i|^2$ but with the random initial phase $\phi_i$ (i.e., $a_i \to a_i e^{i\phi_i}$) that is uniformly distributed in $[0, 2\pi]$. Such numerical calculations provide a random ensemble to mimic the long-time average of the dynamics in a microcanonical ensemble [42]. Throughout this work, all observables are calculated via such ensemble averages.

(a)
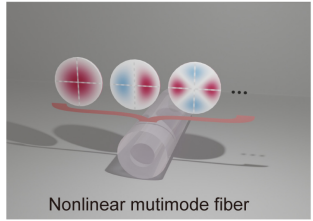
Nonlinear mutimode fiber

(b)
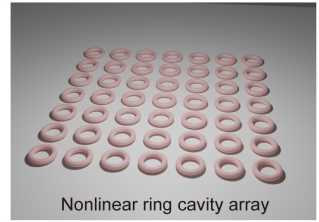
Nonlinear ring cavity array

(c)
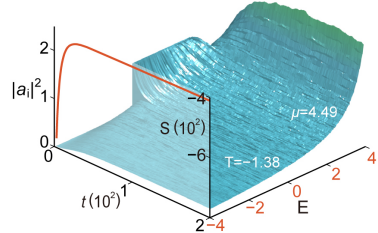

(d)
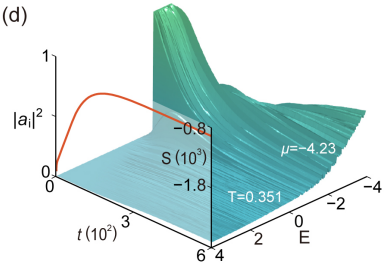

FIG. 1. Emergent photon thermalization in NLMMOSs. (a) Schematic of a coupled ring cavity array with nonlinear effects. (b) Schematic of a nonlinear multi-mode fiber supporting many propagating modes. (c) and (d) Thermalization as revealed by the evolution of the mode occupancy $|a_i|^2$ in a coupled ring cavity array with different initial conditions. We use a simple square lattice array with $20 \times 20$ sites of the same onsite energy. The equilibrium distribution follows the Rayleigh-Jeans form, showing negative and positive temperatures in (c) and (d), separately. The reversed energy axes are labeled by different colors. Red curves give the optical entropy ($S = \sum_{i=1}^{M} \ln|a_i|^2$) as a function of time.

It can be found from Figs. 1(c) and 1(d) that the equilibrium temperature and chemical potential can be well controlled by the initial condition. In Fig. 1(c), we choose the initial distribution as a rectangular distribution with $|a_i|^2 = 1$ for a finite energy region $[0, 3]$. After the thermalization process (for about 40), the system reaches an equilibrium state with $T = -1.38$ and $\mu = 4.49$. If the initial distribution is set as $|a_i|^2 = 1$ for the energy region $[-2, -1]$ as in Fig. 1(d), then the equilibrium state has $T = 0.351$ and $\mu = -4.23$. For these two cases, the conserved internal energy and optical power are $U = 212.10$ and $P = 170$ for Fig. 1(c) ($U = -75.64$ and $P = 51$ for Fig. 1(d)). Figs. 1(c) and 1(d) also show that in the thermalization process, the entropy of the system increases continuously before finally reaching the thermal equilibrium state which is featured by a constant entropy and the Rayleigh–Jeans distribution.

An intriguing property of the thermodynamics in NLMMOSs is that these systems can exhibit negative temperatures. In fact, negative temperature can be obtained by continuously tuning the parameter $\beta = 1/T$ from positive to negative across zero (meanwhile, the temperature goes from positive infinite to negative infinite). Finally, the thermal equilibrium states in NLMMOS can be characterized by the following equation of state [41]:

$$U - \mu P = MT, \qquad (4)$$

which is analogous to the state equation of ideal gas. This equation determines the two intensive quantities $T$ and $\mu$ through the three extensive quantities $M$, $U$, and $P$. Here, it is assumed that the system is completely isolated from the environment. In reality, it is required that the photon lifetime is long enough, which is typically satisfied when the quality factors of the cavities (or waveguides) are $\gtrsim 10^4$—a criterion that can be met in many photonic systems.

**Full counting statistics theory**

In this work, we study the non-equilibrium steady-state transport [61-63] between optical systems, through the method based on the joint particle and energy FCS [64-68]. The core of this theory is to obtain the cumulant generating function (CGF) [69] which contains the full information of the transport current and their fluctuations. To rationalize the derivation process, we need to introduce the two-point measurement scheme.

Consider an observable $\hat{B}(t)$ in the Schrödinger picture, where the time variation of operators only stems from an external driving. The observable $\hat{B}(t)$ will be either an energy operator $\hat{H}$ or a particle number operator $\hat{N}$. The eigenvalues (eigenvectors) of $\hat{B}(t)$ are labeled by $b_t$ ($|b_t\rangle$) and $\hat{B}(t) = \sum|b_t\rangle b_t \langle b_t|$. The two-point measurement scheme contains the measurement with outcome $b_0$ at time $t = 0$ and the measurement with outcome $b_t$ at time $t$ as described by the joint probability

$$P[b_t, b_0] \equiv \mathrm{Tr}\{\hat{P}_{b_t}\hat{U}(t,0)\hat{P}_{b_0}\hat{\rho}_0\hat{P}_{b_0}\hat{U}^\dagger(t,0)\hat{P}_{b_t}\} = P^*[b_t, b_0], \qquad (5)$$

where, $\hat{U}(t,0)$ is the time evolution operator, $\hat{\rho}_0$ denotes the initial density matrix, and the density matrix after time evolution is $\hat{\rho}(t) = \hat{U}(t,0)\hat{\rho}_0\hat{U}^\dagger(t,0)$. The projection operator is $\hat{P}_{b_t} = |b_t\rangle\langle b_t|$, which satisfies the relations $\hat{P}_{b_t} = \hat{P}_{b_t}^2$ and $\sum_{b_t}\hat{P}_{b_t} = \hat{1}$. It is easy to find the normalization of the probabilities, $\sum_{b_0,b_t} P[b_t,b_0] = 1$.

The probability for the difference $\Delta b = b_t - b_0$ between the two measured outcomes is

$$p(\Delta b) = \sum_{b_t,b_0} \delta(\Delta b - b_t + b_0) P[b_t,b_0]. \tag{6}$$

The generating function (GF) for such probability is

$$G(\lambda) \equiv \int_{-\infty}^{\infty} d\Delta b \, e^{i\lambda\Delta b} p(\Delta b) = G^*(-\lambda) = \sum_{b_t,b_0} e^{i\lambda(b_t-b_0)} P[b_t,b_0]. \tag{7}$$

Here, $\lambda$ is the counting field conjugated to the random variable $\Delta b$. By substituting Eq. (5) into Eq. (7), the GF has another expression:

$$G(\lambda) = \text{Tr}\hat{\rho}(\lambda,t). \tag{8}$$

where, $\hat{\rho}(\lambda,t) \equiv \hat{U}_{\lambda/2}(t,0)\hat{\rho}_0\hat{U}^\dagger_{-\lambda/2}(t,0)$ and $\hat{U}_\lambda(t,0) \equiv e^{i\lambda\hat{B}(t)}\hat{U}(t,0)e^{-i\lambda\hat{B}(0)}$. For $\lambda = 0$, $\hat{\rho}(\lambda,t)$ reduces to the system's density matrix $\hat{\rho}(t)$ and $\hat{U}_\lambda(t,0)$ reduces to the standard time evolution operator $\hat{U}(t,0)$.

Taking the logarithm of the GF, we get the expression for the cumulant GF (CGF),

$$\mathcal{Z}(\lambda) = \ln G(\lambda), \tag{9}$$

whose $n^{\text{th}}$ derivative with respect to $\lambda$ evaluated at $\lambda = 0$ gives the $n^{\text{th}}$ cumulant $K_n$ of $p(\Delta b)$.

$$K_n = (-i)^n \frac{\partial^n}{\partial \lambda^n} \mathcal{Z}(\lambda)\bigg|_{\lambda=0}. \tag{10}$$

For instance, the first cumulant $K_1$ gives the average of $\Delta b$, $K_1 = \langle\Delta b\rangle$ which characterizes the energy or particle number change in a thermodynamic system in the time duration *t*. This quantity is then connected to the energy or particle current flowing into the system. $K_2 = \langle\Delta b^2\rangle - \langle\Delta b\rangle^2$ gives the variance of $\Delta b$, and $K_3 = \langle(\Delta b - \langle\Delta b\rangle)^3\rangle$ gives the skewness of $\Delta b$. These and the other higher cumulants give the fluctuations of the energy or particle current.

At nonequilibrium steady states, the cumulants vary linearly with time. Therefore, it is natural to define the long-time limit of the CGF,

$$S(\lambda) = \lim_{t\to\infty} \frac{1}{t}\mathcal{Z}(\lambda). \tag{11}$$

Similar to the CGF, taking derivation on the long-time limit of CGF can directly obtain the long-time average of the above cumulants and give direct information of the transport energy and particle currents as well as their fluctuations.

## Optical thermoelectric transport

In electronic systems with temperature or chemical potential bias, electrons above and below the chemical potential flow according to these biases, leading to net heat and charge currents. This phenomenon is called the thermoelectric effect [70-73]. Similarly, for the transport between two connected NLMMOSs, there can be particle and heat currents, which is a feature distinct from the conventional black body thermal radiation. This feature emerges because the photonic distributions here are characterized by the temperature and the chemical potential---both are tunable. The particle and heat currents should be considered independently in generic transport setups. We term such transport with both particle and heat currents as optical thermoelectric transport and the related effects as the optical thermoelectric effects.

A prototype of such transport systems is based on two infinite photonic reservoirs maintained at different temperatures and chemical potentials, contacting via an embedded small system, as illustrated in Fig. 2(a) (called the two-reservoir setup). The Hamiltonian of the whole system reads $\hat{H} = \hat{H}_{Res} + \hat{H}_S + \hat{V}$, where $\hat{H}_{Res} = \hat{H}_L + \hat{H}_R$ represents the Hamiltonian of the two photonic reservoirs ($L$ and $R$), $\hat{H}_S$ gives the Hamiltonian of the embedded small system, and $\hat{V}$ stands for the interactions among them.

For the total system without external driving, the observable $\hat{B}(t)$ is time-independent. The dynamics of the modified time evolution operator satisfies

$$\frac{d}{dt}\hat{U}_\lambda(t,0) = -i\hat{H}_\lambda(t)\hat{U}_\lambda(t,0), \tag{12}$$

where $\hat{H}_\lambda(t)$ is the modified Hamiltonian defined as

$$\hat{H}_\lambda(t) \equiv e^{i\lambda\hat{B}(t)}\hat{H}(t)e^{-i\lambda\hat{B}(t)}. \tag{13}$$

Further on, the modified evolution operator can be expressed as

$$\hat{U}_\lambda(t,0) = \exp_+\left\{-i\int_0^t d\tau \hat{H}_\lambda(\tau)\right\}$$
$$\hat{U}^\dagger_{-\lambda}(t,0) = \exp_-\left\{i\int_0^t d\tau \hat{H}_{-\lambda}(\tau)\right\}, \tag{14}$$

where the subscripts on the exponents represent different contour-ordering of operators. Given that all the operators mentioned above are in fact time-independent, the density matrix can be simplified as

$$\hat{\rho}(\lambda,t) \equiv e^{-i\hat{H}_\lambda t}\hat{\rho}_0 e^{i\hat{H}_{-\lambda}t}. \tag{15}$$

Here, the observables are chosen as the particle number and energy of the left reservoir, the modified Hamiltonian takes the following form:

$$\hat{H}_\lambda = e^{i/2(\lambda_e\hat{H}_L+\lambda_p\hat{N}_L)}\hat{H}e^{-i/2(\lambda_e\hat{H}_L+\lambda_p\hat{N}_L)} = \hat{H}_0 + \hat{V}_\lambda. \tag{16}$$

Here, $\hat{H}_0 = \hat{H}_{Res} + \hat{H}_S$ is the $\lambda$ independent Hamiltonian, and $\hat{V}_\lambda$ is the modified interaction.

The initial state of the total system is considered as a product state $\rho(0) = \rho_L(0) \otimes \rho_R(0) \otimes \rho_S(0)$. As a result, Eq. (8) can be rewritten in terms of the system's reduced density matrix.

$$G(\lambda) = \text{Tr}_S\text{Tr}_{Res}(\hat{\rho}(\lambda,t)) = \text{Tr}_S\hat{\rho}_S(\lambda,t). \tag{17}$$

We can directly obtain the GF by tracing the small system's density matrix.

Further on, we try to derive the explicit expression of quantum master equation for photon transport based on the energy level presentation. The total system we considered is composed by two photonic reservoirs connected through a two-level system. Notice that, the two reservoirs are decoupled, the only way to transfer photons is using the two-level system. The eigenstates of reservoirs and system are labeled by $i$ and $s$, respectively. The Hamiltonian of the whole transport system is $\hat{H} = \hat{H}_L + \hat{H}_R + \hat{H}_S + \hat{V}$.

$$\hat{H}_X = \sum_{i\in X=L,R}\epsilon_i\hat{c}_i^\dagger\hat{c}_i \quad \hat{H}_S = \sum_{s\in S}\epsilon_s\hat{c}_s^\dagger\hat{c}_s \tag{18}$$

where $\hat{c}^\dagger$ and $\hat{c}$ are the photon creation and annihilation operators, which obey the bosonic commutation relations $[\hat{c}_i,\hat{c}_j^\dagger] = \delta_{ij}$. The interaction between the reservoir and the small system in the center is described by the hopping between the states $s$ and $i$

$$\hat{V}_X = \hat{J}_X^\dagger + \hat{J}_X, \quad \hat{J}_X = \sum_{s\in S, i\in X=L,R} J_s^X \hat{c}_s^\dagger \hat{c}_i, \tag{19}$$

where $J_s^X$ represents the couplings between the reservoir and the eigenmode $s$ of the central small system. In this work, we study the counting statistics of the energy and particle in the left reservoir, which are represented by the Hamiltonian operator and particle number operator, respectively. Therefore, the modified interaction $\hat{V}_\lambda$ in Eq. (16) takes the following form,

$$\hat{V}_\lambda = e^{(i/2)(\lambda_p \hat{N}_L + \lambda_e \hat{H}_L)}(\hat{J}_L + \hat{J}_L^\dagger)e^{(-i/2)(\lambda_p \hat{N}_L + \lambda_e \hat{H}_L)} + \hat{V}_R = \hat{J}_L(\lambda) + \hat{J}_L^\dagger(\lambda) + \hat{V}_R. \tag{20}$$

Using the fact for boson, $\hat{c}_i \hat{H}_L = (\epsilon_i + \hat{H}_L)\hat{c}_i$, $\hat{J}_L(\lambda)$ in Eq. (20) can be expressed as

$$\hat{J}_L(\lambda) = \sum_{s\in S, i\in L} J_s^L \hat{c}_s^\dagger \hat{c}_i e^{(i/2)\lambda_p + (i/2)\lambda_e \epsilon_i}.$$

Further on, we can define two rates correspond to the "up" and "down" jumps between the small system's states,

$$k_u^X = 2\pi g_X(\epsilon_s)|J_s^X|^2[1 + n_X(\epsilon_s)],$$
$$k_d^X = 2\pi g_X(\epsilon_s)|J_s^X|^2 n_X(\epsilon_s),$$
$$k_u = k_u^L + k_u^R, \quad k_d = k_d^L + k_d^R.$$

Here, $g_X(\epsilon_s)$ and $n_X(\epsilon_s)$ are the density of states and the equilibrium distribution of the photonic reservoir $X(=L,R)$ at the energy $\epsilon_s$, respectively. With these notations, the dynamics of the small system's density matrix can be expressed as (see details of the derivation in Appendix C),

$$\dot{\hat{\rho}}_S(\lambda, t) = -i[\hat{H}_S, \hat{\rho}_S(\lambda, t)] + \sum_s -k_u \hat{c}_s^\dagger \hat{c}_s \hat{\rho}_S(\lambda, t) - k_d \hat{c}_s \hat{c}_s^\dagger \hat{\rho}_S(\lambda, t)$$
$$+ (k_d^L e^{i\lambda_p + i\lambda_e \epsilon_i} + k_d^R)\hat{c}_s^\dagger \hat{\rho}_S(\lambda, t)\hat{c}_s$$
$$+ (k_u^L e^{-i\lambda_p - i\lambda_e \epsilon_i} + k_u^R)\hat{c}_s \hat{\rho}_S(\lambda, t)\hat{c}_s^\dagger. \tag{21}$$

Throughout this work, the dot above a symbol stands for the time derivative. In the small system's eigenstates basis $\{\sum_s |n_s\rangle\}$, Eq. (21) describes the population dynamics of the small system.

$$\dot{\rho}_{n_s}(\lambda, t) = \sum_s -k_u n_s \rho_{n_s}(\lambda, t) - k_d(1 + n_s)\rho_{n_s}(\lambda, t) + [k_d^L e^{i\lambda_p + i\lambda_e \epsilon_i} + k_d^R]n_s \rho_{n_s-1}(\lambda, t)$$
$$+ [k_u^L e^{-i\lambda_p - i\lambda_e \epsilon_i} + k_u^R](n_s + 1)\rho_{n_s+1}(\lambda, t). \tag{22}$$

Here, $\dot{\rho}_{n_s}(\lambda, t) \equiv \langle n_s|\dot{\hat{\rho}}_S(\lambda, t)|n_s\rangle$, and the subscript $s$ labels the two eigenstates 1 and 2 of the small system (see Fig. 2(a)). The states of the two-level bosonic system can be denoted as $|n_1, n_2\rangle$ with $n_1, n_2 = 0, 1, 2, \ldots$. The adjacent states $|n_1 \pm 1, n_2\rangle$ and $|n_1, n_2 \pm 1\rangle$, can be connected with $|n_1, n_2\rangle$ by transferring a photon between the small system and the reservoirs. The population dynamics in Eq. (22) can, in fact, be separated into two decoupled parts which can be cast into the matrix form $\dot{\rho}_{n_{1/2}} = M_{1/2} \rho_{n_{1/2}}$.

Unlike the fermions obeying the Pauli exclusion principle, the number of photons in a level is unrestricted. Therefore, the matrix $M_{1/2}$ is infinitely large. In fact, $\rho_{n_{1/2}}$ is an infinite dimensional row vector and $M_{1/2}$ is an infinite tridiagonal matrix,

$$M_{1/2} = \begin{pmatrix} O_2 & U & \cdots & \cdots & \cdots & \cdots & \cdots \\ D & O_1 + 2O_2 & 2U & \cdots & \cdots & \cdots & \cdots \\ \vdots & 2D & 2O_1 + 3O_2 & 3U & \cdots & \cdots & \cdots \\ \vdots & \vdots & 3D & 3O_1 + 4O_2 & 4U & \cdots & \cdots \\ \vdots & \vdots & \vdots & 4D & 4O_1 + 5O_2 & 5U & \cdots \\ \vdots & \vdots & \vdots & \vdots & 5D & 5O_1 + 6O_2 & \cdots \\ \vdots & \vdots & \vdots & \vdots & \vdots & \vdots & \ddots \end{pmatrix}. \tag{23}$$

Here,

$$O_1 = -k_d^L - k_d^R, \quad O_2 = -k_u^L - k_u^R,$$

$$D = e^{i\lambda_p + i\lambda_e \epsilon_i} k_d^L + k_d^R, \qquad U = e^{-i\lambda_p - i\lambda_e \epsilon_i} k_u^L + k_u^R. \tag{24}$$

We find that by truncating the infinite matrix $M_{1/2}$ into a $12 \times 12$ tridiagonal matrix is enough to converge the calculation (see Appendix D for details). In the calculation, we obtain the same number of negative and positive eigenvalues. The long-time limit behavior of the CGF is dominated by the largest negative eigenvalue,

$$S(\lambda) = \lim_{t \to \infty} \frac{1}{t} \ln G(\lambda, t) = \sum_s v_{max}^s = v_{max}^1 + v_{max}^2 \tag{25}$$

By taking derivation of $S(\lambda)$ with respect to the counting field $\lambda$ at $\lambda = 0$, as Eq. (10) states, we can obtain the long-time averaged transport currents and their fluctuation. Although these currents are defined as the currents flowing into the left reservoir, due to the particle and energy conservation, they in fact give information on the currents transported across the small system.

The first order terms in Eq. (10) give the photon current and the energy current,

$$\langle I \rangle = \sum_s \frac{\gamma_s^L \gamma_s^R}{\gamma_s^L + \gamma_s^R} \times [n_L(\epsilon_s) - n_R(\epsilon_s)], \tag{26A}$$

$$\langle I_E \rangle = \sum_s \epsilon_s \frac{\gamma_s^L \gamma_s^R}{\gamma_s^L + \gamma_s^R} \times [n_L(\epsilon_s) - n_R(\epsilon_s)], \tag{26B}$$

where $\gamma_s^X = 2\pi g_X(\epsilon_s)|J_s^X|^2$ gives the transition rate for the small system's eigenmode $s$ to decay into the photonic reservoir $X(= L, R)$.

**Photon transport between two reservoirs**
**A. Model setup**

When the size is considerably large, a NLMMOS can serve as a photon reservoir owing to its tunable thermodynamic properties that completely depend on the initial conditions. To consider the transport effects, we use two large NLMMOSs as the left and right photonic reservoirs. In our system, each reservoir is a square lattice with $10 \times 10$ sites with hopping $\kappa = 1$ and nonlinearity $\chi^{(3)} = \frac{1}{2}$ due to Kerr effects. Therefore, the entire system has two photonic reservoirs and a small system consisting of two sites that bridge the two photonic reservoirs. The Hamiltonian of the whole system is $\widehat{H} = \widehat{H}_L + \widehat{H}_R + \widehat{H}_S + \widehat{V}$, where

$$\widehat{H}_{L/R} = \sum_{<i,j> \in L/R} \kappa \hat{c}_i^\dagger \hat{c}_j + \sum_{i \in L/R} \chi^{(3)} \hat{c}_i^\dagger \hat{c}_i \hat{c}_i^\dagger \hat{c}_i$$

$$\widehat{H}_S = \kappa' \hat{c}_l^\dagger \hat{c}_r + \kappa' \hat{c}_r^\dagger \hat{c}_l \qquad \widehat{V} = \kappa' \hat{c}_l^\dagger \hat{c}_m + \kappa' \hat{c}_r^\dagger \hat{c}_n + H.c. \tag{27}$$

Here, the reservoirs' Hamiltonians, $\widehat{H}_L$ and $\widehat{H}_R$, are the same as in Eq. (2). The central small system consists of two sites ($l$ and $r$) coupled by $\kappa'$. The interaction $\widehat{V}$ describes the couplings between the small system and the left/right reservoirs. $m$ and $n$ denotes the site number in the left and right square lattices that couples with the $l$ and $r$ site, respectively.

Similar to Eqs. (18) and (19), Hamiltonians in Eq. (27) can be written in a dragonized form. For example, the diagonalization of $\widehat{H}_S$ is.

$$\hat{c}_1 = \frac{\sqrt{2}}{2} \hat{c}_l + \frac{\sqrt{2}}{2} \hat{c}_r \qquad \hat{c}_2 = \frac{\sqrt{2}}{2} \hat{c}_l - \frac{\sqrt{2}}{2} \hat{c}_r$$

$$\epsilon_1 = \kappa' \qquad \epsilon_2 = -\kappa'$$

$$\widehat{H}_S = \epsilon_1 \hat{c}_1^\dagger \hat{c}_1 + \epsilon_2 \hat{c}_2^\dagger \hat{c}_2 \tag{28}$$

From the above expressions, the transition rate $\gamma_s^{L(R)}$ can be determined by the Fermi Golden rule:

$$\gamma_{1/2}^{L(R)} = \frac{1}{2} \times 2\pi |\kappa'|^2 g_{1/2}^{m(n)} = \frac{1}{2} \times 2\pi |\kappa'|^2 \left\{ \frac{1}{\pi} \sum_{i \in L(R)} \frac{\Gamma}{\left(\epsilon_i - \epsilon_{1/2}\right)^2 + \Gamma^2} \right\} \left| \psi_{i \in L(R)}^{m(n)} \right|^2. \quad (29)$$

Here, the prepositive $\frac{1}{2}$ stems from the fact that $\left|\psi_{1/2}^{l(r)}\right|^2 = \frac{1}{2}$ where the superscripts $l$ and $r$ present the two sites of the embedded system, and subscripts 1 and 2 denote the two eigenstates of the small system. Both wavefunctions $\psi_{1/2}^{l(r)}$ and $\psi_{i \in L(R)}^{m(n)}$ are normalized at the two-point system and the square lattice, respectively. Therefore, their moduli squares represent the local density of states at positions $l$, $r$, $m$ and $n$. The complex term in the $\{\cdots\}$ of Eq. (29) is the Lorentz function that describes the spectral broadening of the states $\epsilon_1$ and $\epsilon_2$ of the small system.

In our simulations, to reduce the run time and give more information on the transport processes, the initial distribution in each reservoir is set as the equilibrium Rayleigh-Jeans distribution with a randomized phase (uniformly distributed in the region $[0, 2\pi]$) at each eigenstate. Repetitions of the evolution process are set as 3000 times to obtain sufficient data for statistical analysis.

### B. Photon transport at negative temperatures

In this section, we present the non-equilibrium steady-state photon transport in our system as characterized by the long-time limit CGF [Eq. (24)] and compare the results from the FCS theory with the first-principle simulations based on the nonlinear Schrödinger equation. We are more interested in the regime where at least one of the reservoirs has negative temperature. The non-equilibrium steady-state photon transport demands that the distributions in the left and right reservoirs are nearly invariant during the transport. Therefore, the coupling $\kappa'$ connecting the small system to the reservoirs must be set to be very small in the simulations. In addition, the evolution time span in a single simulation is chosen as 10000 time-units to obtain the long-time behaviors of the nonequilibrium transport. Note that as we work with the dimensionless nonlinear Schrödinger equation [Eq. (3)], every quantity has a natural unit (e.g., the energy unit is the hopping coupling in the reservoirs, from which the time unit can be determined via the uncertainty relation; when compared with experiments, the energy and time units can be determined rigorously; see Appendix B for more information).

We now use an example to demonstrate the steady-state transport in our system via the first-principle simulations. As shown in Fig. 2(b), the optical power and the internal energy of the left and right photonic reservoirs change very slowly with time. Although this feature is distinct from the steady-state transport between two reservoirs in the thermodynamic limit, the transport here can still be considered approximately as the steady-state transport. There are other interesting features in Fig. 2(b). We find that the total optical power $P_L + P_R$ is strictly conserved, whereas due to the nonlinear interactions, the total internal energy $U_L + U_R$ (as is defined solely on the linear energy) is not a conserved quantity. Nevertheless, the total internal energy $U_L + U_R$ is approximately conserved at long-time scales, with fast fluctuations at a short timescale of 8 time-units. Thus, thermodynamics and non-equilibrium steady-state transport theories are still valid at very long-time scales.

Fig. 2(c) verifies the steady-state transport from the aspect of the distributions in the reservoirs. It is seen that the initial and final distributions during the entire time interval of 10000 time-units almost overlap with each other, which indicates that the change of the distributions during the whole transport process is negligible---a notable feature of the steady-state transport.

In this section, we study only the photon particle current and its fluctuations which are well-defined in the simulation and the FCS theory. In the simulation, the currents are obtained via the Heisenberg equation of motion, $\hat{I}_{L/R} = \frac{d}{dt}\hat{N}_{L/R} = i[\hat{H}, \hat{N}_{L/R}]$, where $\hat{N}_{L/R} = \sum_{i \in L/R} \hat{c}_i^\dagger \hat{c}_i$ is the total particle number in the left/right reservoir. For the left and right reservoir, we obtain, respectively, the following photon current operators for the left and right systems,

$$\hat{I}_L = i[\kappa' \hat{c}_m^\dagger \hat{c}_l + \kappa' \hat{c}_l^\dagger \hat{c}_m, \hat{c}_m^\dagger \hat{c}_m] = i\kappa'(\hat{c}_l^\dagger \hat{c}_m - \hat{c}_m^\dagger \hat{c}_l),$$
$$\hat{I}_R = i[\kappa' \hat{c}_n^\dagger \hat{c}_r + \kappa' \hat{c}_r^\dagger \hat{c}_n, \hat{c}_n^\dagger \hat{c}_n] = i\kappa'(\hat{c}_r^\dagger \hat{c}_n - \hat{c}_n^\dagger \hat{c}_r). \tag{30}$$

We extract the instant photon current in each simulation for every 2.5 time-units (totally 10000 time-units for a single simulation) by taking the expectation value of the above operators using the instant photonic wavefunction. After 3000 simulations, we obtain an ensemble of 12 million data points of these currents which is sufficiently large to perform statistical analysis. To avoid confusion, we note that, in the following, the symbols $\langle I \rangle$ and $\langle \Delta I \rangle = \langle I^2 \rangle - \langle I \rangle^2$ stands, respectively, for the average and variance of the currents averaged over an ensemble of 3000 repeated simulations with the randomized initial phase condition at a certain time. In contrast, for the currents and their variance averaged over both the ensemble and the long-time scale, the average symbol, $\langle ... \rangle$, is omitted.

The steady-state transport implies: $I_L = -I_R$ and $\Delta I_L = \Delta I_R$. Thus, the photon current from the left reservoir to the right reservoir is equal to $I = -I_L = I_R$. We thus only present the results for the photon currents flowing into the left reservoir. In Fig. 2(d), we find that the 3000-simulation-averaged photon current fluctuates around its long-time-averaged value (i.e., the black dashed line). The variance of photon currents $\Delta I_L$ shown in Fig. 2(e) is larger than the photon current $I_L$. Thus, the direction of photon currents may reverse in the transient dynamics. Besides, fluctuations are significantly more pronounced at the beginning compared to the rest of the time. This feature indicates that the system reaches steady-state transport at about t=3000. Therefore, we only use the data from t=3000 to t=10000 for the long-time averages.

The photon current and its variance averaged over both the ensemble and the long-time scale are labeled by the black dashed lines in Figs. 2(d) and 2(e). These averaged values are $I_L = -2.57 \times 10^{-4}$ and $\Delta I_L = 0.72 \times 10^{-3}$. Surprisingly, the FCS theory gives comparable values through Eq. (10): $I_L = -2.65 \times 10^{-4}$ and $\Delta I_L = 1.20 \times 10^{-3}$, although these two approaches are quite different. On one hand, the FCS theory assumes that the reservoirs are in the thermodynamic limit, i.e., infinitely large, and are kept at thermal equilibrium. The FCS theory is essentially a stochastic approach based on the quantum perturbation theory. On the other hand, the simulation is based solely on the nonlinear Schrödinger equation, which contains purely deterministic dynamics. The consistency between the two approaches suggests that the statistical theory agrees well with the deterministic dynamic theory in our system, which is quite noteworthy. These results indicate that it is possible to compare statistical physics theory and dynamic simulations quantitatively in certain regimes when thermodynamic behaviors can be captured by the first-principle calculations of the deterministic dynamics.

We further examine the photon current and its variance for various system parameters. In Fig. 2(f), we study the dependence of these quantities on the system-reservoir coupling $\kappa'$. It is found that the photon current increases with $\kappa'$ in nearly quadratic form, while the variance of the photon current also increases with $\kappa'$ in similar form but much more rapidly. The quadratic increase of the photon current is consistent with Eq. (26) since both $\gamma_s^L$ and $\gamma_s^R$ are proportional to $\kappa'^2$. At large

$\kappa'$, higher-order corrections come into play and the dependence deviates from $\sim \kappa'^2$. The calculation results from the first-principle simulations and the FCS theory are comparable and qualitatively consistent with one another. In particular, the photon currents obtained from the two approaches agree excellently with each other. In Fig. 2(g), we investigate the photon current and its variance when only the temperature of the left photonic reservoir $T_L$ changes. It is seen that as $T_L$ increases, the temperature difference increases and the photon current ramps up, while the variance of the photon current also increases. At $T_L = -2$, there is no bias in the temperature or chemical potential. The average photon current vanishes. Nevertheless, the variance of the photon current is nonzero---an intrinsic feature of thermodynamic equilibrium.

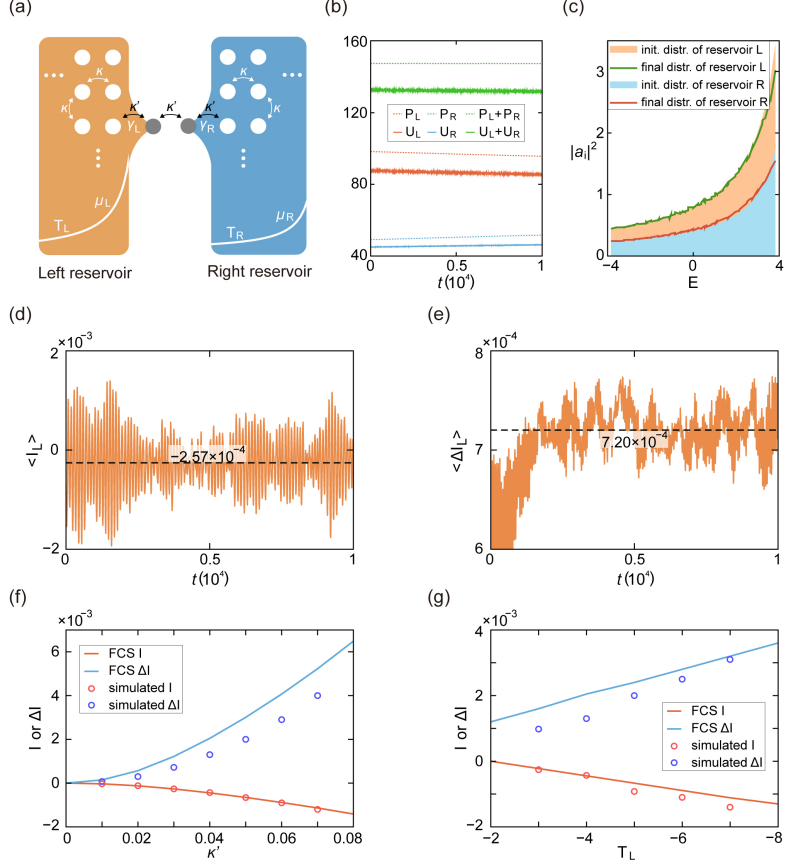

FIG. 2. Transport and fluctuations between two photonic reservoirs. (a) Schematic illustration of the system. Orange and blue regions stand for the left and right reservoirs, respectively. White dots in these regions represent the cavities in the square-lattice nonlinear optical systems that form the two reservoirs. The nearest neighbor coupling between two cavities is $\kappa$. A small system with two cavities (gray dots) serves as the bridge between the two reservoirs. The couplings between the two sites of the small system and every two nearest sites of the individual reservoirs are set as $\kappa'$. $\gamma_L$ and $\gamma_R$ represent the hopping rates of photons transferred from the left and right reservoirs to the small system. (b) The evolution of the optical power $P$ and the internal energy $U$ with time. (c) The initial and final distributions of the two reservoirs from the simulation. (d) The photon current after averaging over 3000 simulations (orange curves) and its long-time averaged value (black dashed line). (e) The variance of the photon current (orange curve) and its long-time averaged value (black dashed line). In (b)-(e), $\kappa' = 0.03$. (f) The dependence of the averaged photon current on the coupling $\kappa'$. (g) The long-time averages and variances of photon currents changing with $T_L$ when $\kappa' = 0.04$. Red and cyan curves represent the results from the FCS (FCS) theory. Red and blue circles denote the results from the simulations based on the nonlinear Schrödinger equation. The parameters for the two reservoirs are $T_L = -4$, $\mu_L = 5$, $T_R = -2$, and $\mu_R = 5$, unless otherwise specified.

## C. Loop currents between positive- and negative-temperature reservoirs

We now study the intriguing transport between positive- and negative-temperature reservoirs and reveal an interesting phenomenon we have named "loop currents.". We stress that the setup studied here is not available in conventional systems. To be concrete, we set the initial conditions for the left and right reservoirs as: $T_L = -2$, $\mu_L = 5$, $T_R = 2$, and $\mu_R = -5$. It is crucial to note that the spectra of the two reservoirs and the small system are symmetric with respect to zero energy (for a specific optical system, the zero energy is defined as the average onsite energy for the cavities). In that case, the distributions in the left and right reservoirs are inverted. Meanwhile, the coupling between the two sites in the small system gives rise to an even-parity eigenstate (anti-bonding state) with positive energy and an odd-parity eigenstate (bonding state) with negative energy. Tunneling via the even state transfers photons from the left reservoir to the right reservoir in the positive energy channel, whereas the odd state offers a negative energy channel to transfer photons from the right reservoir to the left reservoir. These transferred photons are then thermalized via the nonlinear interaction. For instance, high- (low-) energy photons in the right (left) reservoir give (take) energy to (from) other photons in the same reservoir via nonlinear interaction. This scenario, as schematically depicted in Fig. 3(a), gives rise to the phenomenon termed as the loop current: In the positive energy sector, photons flow from the left reservoir to the right, while in the negative energy sector, photons flow from the right reservoir to the left reservoir. In each reservoir, nonlinear interaction enables energy exchange and thermalization [74]. Eventually, the distributions in the two reservoirs become the same, and the whole system reaches thermodynamic equilibrium with infinite temperature (Fig. 3(b)).

In Fig. 3(b), we give the distributions in the two reservoirs at different times, $t = 0$, 500, and 2500 from the first-principle simulations (To enlarge the loop current effect, we set $\kappa' = 1$). These snapshots indicate the evolution of the distributions in the reservoirs. Initially, due to the inversed temperatures and chemical potentials, the distributions in the two reservoirs are related to each other by the particle-hole symmetry, i.e., mirror symmetric with respect to the zero energy. At $t = 500$, this symmetry still holds, indicating that photons with opposite energies are transported at the same rate, which agrees with the scenario illustrated in Fig. 3(a). After a long time, at $t = 2500$, the distributions in the two reservoirs become flat and identical, indicating the equilibrium state of the whole system which has infinite temperature. It is worth noting that the photon distribution in this work is only proportional to the actual photon numbers (up to a constant determined by the total optical power; see Appendix B for the relation between the dimensionless quantities and physical quantities). In fact, it is difficult to determine the actual photon number in experiments, while the distribution can be measured easily up to a constant determined by the photon detection probability.

To reveal the loop current effect, we present the evolution of the currents in the positive and negative energy sectors in Fig. 3(c). It is found that these two currents are approximately opposite to each other. By adding these two currents, we find that the net current is fluctuating around zero in Fig. 3(d). In fact, the long-time averaged net current is close to zero, confirming again the loop current picture in Fig. 3(a). To observe non-zero net photon current, we need to break the particle-hole symmetry. This can be done by raising the energy levels of the small system by 1, as in Fig.

3(e). With this change, we find from simulations that a finite net photon current emerges near the initial time.

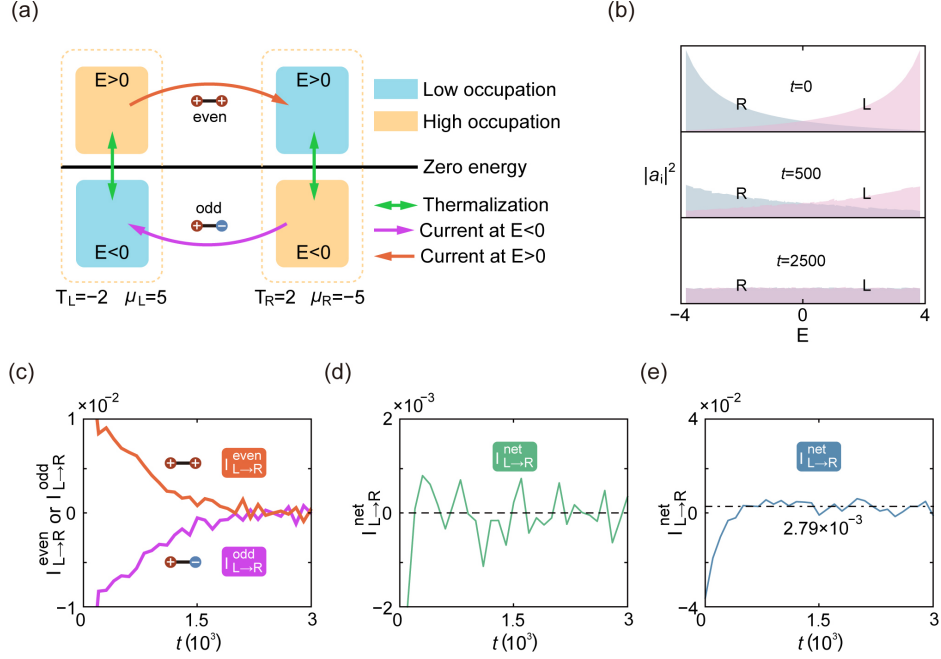

FIG. 3. Transport between two reservoirs with opposite temperatures and chemical potentials. (a) Schematic of the loop current effect when the system has particle-hole symmetry. Red and purple arrows denote the opposite currents in the positive and negative energy sectors via, respectively, the even and odd modes of the small system. Green arrows denote the energy exchange between the positive and negative energy sectors in each reservoir due to thermalization processes via nonlinear interactions. (b) Distributions in the two reservoirs at three different times: $t = 0$, 500 and 2500. (c) Evolution of photon current in the positive and negative energy sectors. Current in the positive (negative) energy sector is through the even (odd) mode of the small system which is denoted by the superscript 'even' ('odd'). (d) Evolution of the net photon current for the same condition as in (c). The black dashed line indicates the zero current around which the net current fluctuates. (e) Evolution of the net photon current when the onsite energy of the two cavities in the small system is raised from 0 to 1. The number gives the long-time averaged (starting from $t = 1250$) value of the current which is labeled by the black dot-dashed line. The parameters are $\kappa' = 1$, $T_L = -2$, $\mu_L = 5$, $T_R = 2$, and $\mu_R = -5$. For all calculations, the current is averaged in the time interval of 100 time-units. To keep the particle-hole symmetry of the whole system, the sign of the nonlinear coefficient $\chi^{(3)}$ is flipped for adjacent sites in each reservoir, while its amplitude is fixed to 1/2.

## Heat exchange fluctuation theorem and the thermodynamic uncertainty relation

In stochastic thermodynamics, the fluctuation theorem [75-78] plays a crucial role. Using step-by-step coarse-graining procedures, the fluctuation theorem can be derived from the detailed balance, which arises from micro-reversibility [79]. The common fluctuation theorems include the Jarzynski's equality for work [80-82], the exchange fluctuation theorem for heat and matter [50, 51], and the fluctuation theorem for entropy production [83, 84].

Previous studies on the fluctuation theorem focus on the positive temperature regime. Here, we try to verify the heat exchange fluctuation theorem at negative temperatures in NLMMOS. The basic setup of our system is two coupled reservoirs (i.e., two square lattices with $10 \times 10$ sites) which are coupled via optical cross phase modulation (XPM) interactions, as illustrated in Fig. 4(a). Unlike in the previous section, here there is no small system acting as the bridge between the two reservoirs. Instead, the two reservoirs are coupled by several pairs of sites. Within each pair, the two sites, one from each reservoir, are interconnected through XPM. In this case, the XPM allows energy exchange between the two reservoirs but no photon transfer between the two reservoirs. Initially, the left reservoir has a Rayleigh-Jeans distribution with $T_L = -3$ and $\mu_L = 5$, while the right reservoir has $T_R = -2$ and $\mu_R = 5$. In the simulation, we randomly establish the XPM coupling on three pairs of sites (both these sites and their connections are randomly chosen). Here, the nonlinear coefficient $\chi^{(3)}$ for the XPM coupling and the Kerr effect are set as $\frac{1}{8}$ and $\frac{1}{4}$, respectively. The XPM coupling is crucial for the energy exchange between the two reservoirs. Therefore, the internal energy in the two reservoirs, $U_L$ and $U_R$, evolve with time, whereas the optical powers, $P_L$ and $P_R$, do not change with time since there is no photon exchange between the two reservoirs (see Fig. 3(b)).

In our two-reservoir system, the heat exchange fluctuation theorem takes the form:
$$\langle e^{-(\beta_L - \beta_R)Q_L} \rangle = 1, \tag{31}$$
where $Q_L$ stands for the heat change of the left reservoir between two measured times. Here, $\langle \cdots \rangle$ represents the ensemble average. $\beta_L$ and $\beta_R$ ($T_L$ and $T_R$) evolve with time due to energy exchange and the finite size of the reservoirs, as shown in Fig. 4(c). Since our two-reservoir system is isolated (without external drive) and photon transport is prohibited (without internal work: $W = \mu \Delta P$), according to the first law of thermodynamics $\Delta U = W + Q$, the heat change $Q_L$ can be simply obtained from the internal energy change $\Delta U_L$.

Numerically, we perform the simulation with the same initial condition (but randomized phases on $a_i$ and random XPM pairs) for 3000 times. These 3000 simulations are used as an ensemble to study the statistics of the heat exchange. Figure 4(d) presents the three quantities, $\langle (\beta_L - \beta_R)Q_L \rangle$, $\sigma^2/2$ ($\sigma^2$ means the variance of $(\beta_L - \beta_R)Q_L$), and $\langle e^{-(\beta_L - \beta_R)Q_L} \rangle$, for different time spans $\Delta t = 2500, 3750$, and $5000$ with the same initial time $t = 5000$. The quantity $\beta_L - \beta_R$ is approximated as the average value in the corresponding time span, i.e., $\beta_L - \beta_R = 0.138, 0.136$, and $0.133$, separately. In Fig. 4(d), we show the values of $\langle (\beta_L - \beta_R)Q_L \rangle$ and $\sigma^2/2$ for the three time-spans and find that these values agree well with each other. In addition, the ensemble average $\langle e^{-(\beta_L - \beta_R)Q_L} \rangle$ is very close to 1. These results verify numerically Eq. (31). We further give the histogram distributions of $\langle (\beta_L - \beta_R)Q_L \rangle$ from numerical calculations for the three time-spans in Fig. 3(e). It is found that these distributions are in excellent agreement with Gaussian profile in linear response regions. With this numerical evidence, we verify the fluctuation theorem in negative temperature regime, thus extending the fluctuation theorem to an unconventional realm.

Recently, thermodynamic uncertainty relations have been proposed and studied for classical and quantum steady states transport, revealing the intriguing trade-off relation between the relative current fluctuations and the dissipation. For instance, a thermodynamic uncertainty relation restricts the average accumulated current $\langle Q \rangle$, its variance $var(Q)$, and the entropy production $\Delta S$ in a nonequilibrium process as follows [85-88],
$$\frac{var(Q)}{\langle Q \rangle^2} \Delta S \geq 2, \tag{32}$$
where we have set $k_B = 1$. To quantify noises and fluctuations in the nonlinear photonic system studied here, we verify whether the thermodynamic uncertainty relation holds in our system. Based on the same setup as in Fig. 4(a)-(e), we quantify the ratio $\frac{var(Q_R)}{\langle Q_R \rangle^2} \Delta S_R$ (subscripts here denote the reservoir) for the reservoir $R$ from first-principle simulations and present the results for different values of $T_L$ in Fig. 4(f) as the solid lines, whereas the lower bound in Eq. (32) is presented as the black dashed line. To investigate the problem carefully, we present the results for different time spans of the nonequilibrium transport process. From Fig. 4(f), we find that the thermodynamic uncertainty relation also holds for negative temperature reservoirs and for nonequilibrium processes involving such reservoirs. This remarkable finding generalizes the previous understanding and investigation of the thermodynamic uncertainty relation from the positive temperature regime to the negative temperature regime. This finding underscores that the polarity of the reservoir temperature, whether it is positive or negative, does not disrupt the thermodynamic uncertainty relation. In a sense, our results strengthen the physical understanding that for systems with a finite spectral range, both positive temperatures and negative temperatures have well-defined thermodynamic properties. In addition, it is noteworthy that as the measurement time interval $\Delta t$ increases, the thermodynamic uncertainty at different temperatures gradually decreases toward the lower bound 2. This tendency is attributed to the enhanced accuracy of the measured thermodynamic currents with prolonged measurement time which leads the system towards reduced uncertainty.

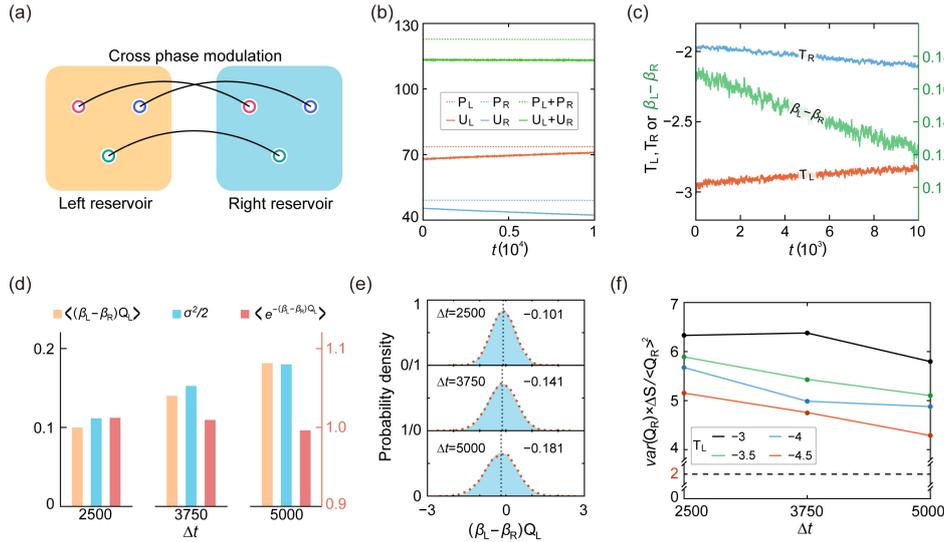

FIG. 4. Heat exchange between negative temperature reservoirs and the fluctuation theorem. (a) Illustration of the system with heat exchange between two reservoirs that are coupled via the XPM nonlinear interaction (indicated by the three black curves). (b) Evolution of the optical power in the left and right reservoirs, $P_L$ and $P_R$, and their summation as well as the internal energy in the left and right reservoirs, $U_L$ and $U_R$, and their summation. The total particle number and internal energy are conserved. (c) Evolution of the temperatures of the left and right reservoirs and the difference in the inverse temperature for the two reservoirs $\beta_L - \beta_R$ (its value is given by the right vertical axis). (d) Three quantities $\langle(\beta_L - \beta_R)Q_L\rangle$, $\sigma^2/2$, and $\langle \exp(-(\beta_L - \beta_R)Q_L)\rangle$, measured during different time spans $\Delta t$ (Here $\sigma^2$ is the variance of $(\beta_L - \beta_R)Q_L$). From left to right, the values of the colored columns are, respectively, {0.101,0.113,1.01}, {0.141,0.153,1.01}, and {0.181,0.180,1.00} for those three quantities. Note that the value of $\langle(\beta_L - \beta_R)Q_L\rangle$ is given by the vertical axis at the right side. (e) Normalized probability density distributions of $(\beta_L - \beta_R)Q_L$ calculated for different time spans $\Delta t$ (red dotted lines) which are close to Gaussian distributions (blue-shaded region) with the same average value and variance of $(\beta_L - \beta_R)Q_L$ for each case. The number at the right side of each subfigure gives the average value $\langle(\beta_L - \beta_R)Q_L\rangle$. (f) The thermodynamic ratio $\frac{var(Q_R)}{\langle Q_R\rangle^2}\Delta S_R$ as a function of the time span $\Delta t$ for different values of $T_L$. Black dashed line represents the lower bound of the thermodynamic uncertainty relation (i.e., 2).

**Conclusion and discussions**

Starting from two distinct approaches, i.e., the first-principle simulations of dynamics from a deterministic nonlinear dynamic equation and the FCS theory based on stochastic processes, we reveal the intriguing transport and fluctuation properties of the NLMMOSs whose equilibrium states are described by the Rayleigh-Jeans distributions. In particular, we explore the steady-state photon transport and its fluctuations between two finite-sized reservoirs. We find that, surprisingly, these two distinct approaches yield quantitatively comparable results, showing a unified picture of deterministic dynamics and statistical physics in the description of nonequilibrium transport in the observed regimes. As the NLMMOSs have finite-ranged photon spectrum, the photon distribution can have negative temperatures---a remarkable property of micro-canonical ensembles. We further reveal that under the circumstance of two reservoirs with opposite temperatures and chemical potentials, an intriguing phenomenon termed as the loop current effect can emerge where the current in the positive energy sector is opposite to that of the negative energy sector. Finally, by quantitatively comparing the statistics of the heat current between the two reservoirs from two distinct approaches, we verify numerically the fluctuation theorem in the negative temperature regime. Our study reveals that the photonic transport properties and their statistical fluctuations in NLMMOSs contain rich and intriguing phenomena which are yet to be explored in experiments. Experimental investigations of these phenomena can gain insight in fundamental statistical physics and motivate future studies such as engineering non-Planckian heat and photon transfer in nonlinear optical fibers and coupled optical resonator systems that may lead to novel discoveries and potential applications in photonics.

## Appendix A: Derivation of the optical entropy

The Gibbs entropy of the system is

$$S = -\int_0^\infty \rho(|a_1|^2,\cdots,|a_M|^2)\ln\rho(|a_1|^2,\cdots,|a_M|^2)\prod_{i=1}^M d|a_i|^2 \tag{A1}$$

where, $\rho$ is the probability density distribution as a function of the mode occupancies $|a_1|^2,\cdots,|a_M|^2$.

Following Kaufman's derivation [1], the optical wave's entropy associated with the $i$th optical mode is

$$s_i = -\int_0^\infty p_i(|a_i|^2)\ln p_i(|a_i|^2)d|a_i|^2. \tag{A2}$$

Here, $p_i(|a_i|^2)$ represents the probability density of finding $|a_i|^2$ photons in the $i$th optical mode. The probability density $p_i(|a_i|^2)$ for the $i$th optical mode obeys the following relation, $\int_0^\infty p_i(|a_i|^2)d|a_i|^2 = 1$. Thus, the expected mode occupancy for the $n$th mode is

$$\langle|a_i|^2\rangle = \int_0^\infty |a_i|^2 p_i(|a_i|^2)d|a_i|^2. \tag{A3}$$

Then, we can obtain the probability density function at thermal equilibrium with maximized $s_n$,

$$p_i(|a_i|^2) = \frac{1}{\langle|a_i|^2\rangle} e^{-\frac{|a_i|^2}{\langle|a_i|^2\rangle}}. \tag{A4}$$

According to the probability equation of independent events, the probability density distribution can be rewritten as:

$$\rho(|a_1|^2,\cdots,|a_M|^2) = \prod_{i=1}^M p_i(|a_i|^2). \tag{A5}$$

Substituting Eq. (A5) into the Gibbs entropy (Eq. (A1)), we can obtain the following expression for the total entropy.

$$S = \sum_{n=1}^M s_i = M + \sum_{i=1}^M \ln(|\langle a_i\rangle|^2). \tag{A6}$$

Maximization of this equation leads to the Rayleigh-Jeans distribution for eigen modes of the system. And the second term in equation works as the relative entropy $\sum_{n=1}^M \ln(|\langle a_n\rangle|^2)$. By reducing the expectation symbol, we can obtain the optical entropy,

$$S = \sum_{i=1}^M \ln|a_i|^2. \tag{A7}$$

## Appendix B: Relation between dimensionless quantities and physical quantities

To simplify the numerical calculations, the Boltzmann constant and the reduced Plank constant are set as 1, giving rise to calculations with dimensionless quantities.

The original Schrödinger equation has the following form:

$$i\hbar\frac{dA_m}{d\tau} + \hbar\omega_m A_m + \kappa_0 \sum_n A_n + \chi_0^{(3)}|A_n|^2 A_n = 0.$$

By taking the transformation $\kappa_0\tau/\hbar = t$ and $A_n\sqrt{\frac{2\chi_0^{(3)}}{\kappa_0}} = a_n$, one can derive the dimensionless Eq. (3) in the main text with $\kappa = 1$ and $\chi^{(3)} = 1/2$. Moreover, we define the energy unit as $E_0 = \kappa_0$, with $\chi_0^{(3)} \sim \mu eV \approx 1.6 \times 10^{-25}$ (J). In this scheme, the time unit is $t_0 = \hbar/E_0$ with $\hbar = 1.05 \times 10^{-34}$ (J·s), and the temperature unit is $T_0 = E_0/k_B$ with $k_B = 1.38 \times 10^{-23}$ (J/K). The constant energy $\hbar\omega_m$ (uniform for all cavities) can be dropped as it has no effect on the physical properties.

In genuine experiments, the optical power varies according to the excitation power of the laser. If the excitation is in the range from 1 $\mu$J to 1nJ, then approximately the optical power is in the range of $10^{10}$ to $10^{13}$. Thus, the actual photon number in each cavity is about $10^8 \sim 10^{11}$ which is much larger than the dimensionless number $|a_i|^2$ in the numerical calculation. In this sense, the physics of photons in NLMMOSs is in the semiclassical regime. The system's thermodynamic behaviors are emergent statistical phenomena that can be controlled by the parameters such as the optical power, the internal energy, the lattice Hamiltonian, and the nonlinear interactions.

## Appendix C: Derivation of the quantum master equation

The density matrix (Eq. (15)) satisfies the equation of motion:
$$\dot{\hat{\rho}}(\lambda,t) = \breve{L}_\lambda \hat{\rho}(\lambda,t) = -i[\hat{H}_\lambda \hat{\rho}(\lambda,t) - \hat{\rho}(\lambda,t)\hat{H}_{-\lambda}] = (\breve{L}_0 + v\breve{L}'_\lambda)\hat{\rho}(\lambda,t)$$
$$= -i[\hat{H}_0, \hat{\rho}(\lambda,t)] - vi[\hat{V}_\lambda \hat{\rho}(\lambda,t) - \hat{\rho}(\lambda,t)\hat{V}_{-\lambda}] \quad (C1)$$

Where $\breve{L}_\lambda$ is a super-operator denoted by a breve and the scalar $v$ multiples $\hat{V}$ is used to label the order in the perturbation expansion below.

In the interaction picture, Eq. (C1) can be simply expressed as
$$\dot{\hat{\rho}}_I(\lambda,t) = v\breve{L}'_\lambda(t)\hat{\rho}_I(\lambda,t) \quad (C2)$$
where $\hat{\rho}_I(\lambda,t) = e^{-\breve{L}_0 t}\hat{\rho}(\lambda,t) = e^{i\hat{H}_0 t}\hat{\rho}(\lambda,t)e^{-i\hat{H}_0 t}$ and $\breve{L}'_\lambda(t) = e^{-\breve{L}_0 t}\breve{L}'_\lambda e^{\breve{L}_0 t}$.

By integrating Eq. (C2) and truncating it to second order ($v^2$), we can obtain the perturbative expansion
$$\hat{\rho}_I(\lambda,t) = \breve{W}(\lambda,t)\hat{\rho}(0) = e^{-\breve{L}_0 t}e^{\breve{L}_\lambda t}\hat{\rho}(0) = [\breve{W}_0(\lambda,t) + v\breve{W}_1(\lambda,t) + v^2\breve{W}_2(\lambda,t) + O(v^3)]\hat{\rho}(0) \quad (C3)$$

where,
$$\breve{W}_0(\lambda,t) = \breve{1}$$
$$\breve{W}_1(\lambda,t) = \int_0^t dT \breve{L}'_\lambda(T)$$
$$\breve{W}_2(\lambda,t) = \int_0^t dT \int_0^T d\tau \breve{L}'_\lambda(T)\breve{L}'_\lambda(T-\tau).$$

The inverse of $\breve{W}(t)$ is
$$\breve{W}^{-1}(\lambda,t) = \breve{W}_0(\lambda,t) - v\breve{W}_1(\lambda,t) + v^2[\breve{W}_1^2(\lambda,t) - \breve{W}_2(\lambda,t)] + O(v^3) \quad (C4)$$
which satisfies $\breve{W}(\lambda,t)\breve{W}^{-1}(\lambda,t) = \breve{1} + O(v^3)$.

For the convenience of the following discussion, we need note that
$$\dot{\breve{W}}(\lambda,t)\breve{A}\breve{W}^{-1}(\lambda,t) = v\dot{\breve{W}}_1(\lambda,t)\breve{A} + v^2[\dot{\breve{W}}_2(\lambda,t)\breve{A} - \dot{\breve{W}}_1(\lambda,t)\breve{A}\breve{W}_1(\lambda,t)] + O(v^3) \quad (C5)$$

Next, using Liouville space notation, the projection super-operator working in reservoir space is defined as
$$\breve{P} = \sum_r |\rho_R^{eq}\rangle\rangle\langle\langle rr|$$

Where $\rho_R^{eq}$ is the equilibrium density matrix of the reservoir. $\check{P}$ and $\check{Q} = \check{1} - \check{P}$ satisfy the normal properties of projection super-operators: $\check{P}^2 = \check{P}$, $\check{Q}^2 = \check{Q}$, and $\check{P}\check{Q} = \check{Q}\check{P} = 0$. After operating on the density matrix, the projection operator is given by

$$\check{P}|\rho(\lambda, t)\rangle\rangle = |\rho_S(\lambda, t)\rangle\rangle \otimes |\rho_R^{eq}\rangle\rangle$$

Now, let $\check{P}$ and $\check{Q}$ act on the total system density matrix in the interaction picture (Eq. (C3)).

$$\check{P}|\rho_I(\lambda, t)\rangle\rangle = \check{P}\widetilde{W}(t)(\check{P} + \check{Q})|\rho_I(0)\rangle\rangle \quad (C6)$$

$$\check{Q}|\rho_I(\lambda, t)\rangle\rangle = \check{Q}\widetilde{W}(t)(\check{P} + \check{Q})|\rho_I(0)\rangle\rangle \quad (C7)$$

Then we assume an initial condition $\check{Q}|\rho(0)\rangle\rangle = 0$, where the reservoir part is diagonal. Using the relation $|\rho_I(0)\rangle\rangle = \widetilde{W}^{-1}(\lambda, t)|\rho_I(\lambda, t)\rangle\rangle$, we can obtain the following results by taking time derivation on Eqs. (C6) and (C7).

$$\check{P}|\dot{\rho}_I(\lambda, t)\rangle\rangle = \check{P}\dot{\widetilde{W}}(\lambda, t)\check{P}\widetilde{W}^{-1}(\lambda, t)\check{P}|\rho_I(\lambda, t)\rangle\rangle + \check{P}\dot{\widetilde{W}}(\lambda, t)\check{P}\widetilde{W}^{-1}(\lambda, t)\check{Q}|\rho_I(\lambda, t)\rangle\rangle \quad (C8)$$

$$\check{Q}|\dot{\rho}_I(\lambda, t)\rangle\rangle = \check{Q}\dot{\widetilde{W}}(\lambda, t)\check{P}\widetilde{W}^{-1}(\lambda, t)\check{P}|\rho_I(\lambda, t)\rangle\rangle + \check{Q}\dot{\widetilde{W}}(\lambda, t)\check{P}\widetilde{W}^{-1}(\lambda, t)\check{Q}|\rho_I(\lambda, t)\rangle\rangle \quad (C9)$$

Using the previous result (Eq. (C5)), we can know that

$$\check{P}\dot{\widetilde{W}}(\lambda, t)\check{P}\widetilde{W}^{-1}(\lambda, t)\check{Q} = v\check{P}\dot{\widetilde{W}}_1(\lambda, t)\check{P}\check{Q} + v^2\check{P}\dot{\widetilde{W}}_2(\lambda, t)\check{P}\check{Q} - v^2\check{P}\dot{\widetilde{W}}_1(\lambda, t)\check{P}\widetilde{W}_1(\lambda, t)\check{Q} + O(v^3)$$

Obviously, the first two terms are zero due to $\check{P}\check{Q} = 1$. And the third term $\check{P}\dot{\widetilde{W}}_1(\lambda, t)\check{P} = \sum_{r,r'} |\rho_R^{eq}\rangle\rangle\langle\langle rr|\check{L}_I'(\lambda, t)|\rho_R^{eq}\rangle\rangle\langle\langle r'r'|$ also vanishes since $[\hat{\rho}_R^{eq}, \widehat{H}_R] = 0$. As a result, by truncating the expansion to $v^2$, the $\check{P}$ projected density matrix evolution seems decoupled from the $\check{Q}$ projected part.

Similarly, using Eq. (C5), the first term in Eq. (C8) can be written as

$$\check{P}\dot{\widetilde{W}}(\lambda, t)\check{P}\widetilde{W}^{-1}(\lambda, t)\check{P} = v\check{P}\dot{\widetilde{W}}_1(\lambda, t)\check{P} + v^2\check{P}\dot{\widetilde{W}}_2(\lambda, t)\check{P} - v^2\check{P}\dot{\widetilde{W}}_1(\lambda, t)\check{P}\widetilde{W}_1(\lambda, t)\check{P} + O(v^3) \quad (C10)$$

The nonzero component in the right side is the second one, and Eq. (C8) can be written as

$$\check{P}|\dot{\rho}_I(\lambda, t)\rangle\rangle = v^2\check{P}\int_0^t d\tau \check{L}_\lambda'(t)\,\check{L}_\lambda'(t-\tau)\check{P}|\rho_I(\lambda, t)\rangle\rangle + O(v^3)$$

Leaving the interaction picture and using the relation $\check{P}e^{-\check{L}_0 t} = e^{-\check{L}_S t}\check{P}$, one can have

$$\check{P}|\dot{\rho}(\lambda, t)\rangle\rangle = \check{L}_S\check{P}|\rho(\lambda, t)\rangle\rangle + v^2 e^{\check{L}_S t}\check{P}\int_0^t d\tau \check{L}_\lambda'(t)\,\check{L}_\lambda'(t-\tau) \times e^{-\check{L}_S t}\check{P}|\rho(\lambda, t)\rangle\rangle \quad (C11)$$

Tracing Eq. (C11), we obtain

$$\dot{\hat{\rho}}(\lambda, t) = \check{L}_S\hat{\rho}_S(\lambda, t) + v^2 \sum_r \int_0^t d\tau e^{\check{L}_S t}\langle\langle rr|\check{L}_\lambda'(t)\,\check{L}_\lambda'(t-\tau) \times |\rho_R^{eq}\rangle\rangle e^{-\check{L}_S t}\check{P}\hat{\rho}_S(\lambda, t) \quad (C12)$$

Finally, by explicitly expressing these super-operators, we can get the evolution of the small system's density matrix, also called general quantum master equation (GMQE),

$$\dot{\hat{\rho}}_S(\lambda, t) = -i[\widehat{H}_S, \hat{\rho}_S(\lambda, t)]$$

$$+ \sum_{ss'} \int_0^t d\tau [-\text{Tr}_{Res}\{\hat{V}_\lambda^s \hat{V}_\lambda^{s'}(-\tau)\hat{\rho}_{Res}^{eq}\hat{\rho}_S(\lambda, t)\} - \text{Tr}_{Res}\{\hat{\rho}_{Res}^{eq}\hat{\rho}_S(\lambda, t)\hat{V}_{-\lambda}^s(-\tau)\hat{V}_{-\lambda}^{s'}\}$$

$$+ \text{Tr}_{Res}\{\hat{V}_\lambda^s \hat{\rho}_{Res}^{eq}\hat{\rho}_S(\lambda, t)\hat{V}_{-\lambda}^{s'}(-\tau)\} + \text{Tr}_{Res}\{\hat{V}_\lambda^s(-\tau)\hat{\rho}_{Res}^{eq}\hat{\rho}_S(\lambda, t)\hat{V}_{-\lambda}^{s'}\}, \quad (C13)$$

where $\hat{V}_\lambda^s(t) = e^{i\widehat{H}_0 t}\hat{V}_\lambda^s e^{-i\widehat{H}_0 t}$ and $\hat{V}_\lambda^s$ represents the modified interaction between the reservoir and the eigenstate $s$ of the system. $\hat{\rho}_{Res}^{eq}$ presents the equilibrium density matrix. By solving this equation, we can obtain the GF which describes the change of observable quantities between two measurements.

For a two-reservoir transport model with an embedded two-level system, its Hamiltonian is $\hat{H} = \hat{H}_L + \hat{H}_R + \hat{H}_S + \hat{V}$. The explicit interaction and sub-Hamiltonians are described in the main text (see Eqs. (18-20)).

Based on the Eq. (20), the modified interaction after time $t$ reads

$$\hat{V}_\lambda(t) = e^{i\hat{H}_0 t}\hat{V}_\lambda e^{-i\hat{H}_0 t}$$

$$= \sum_{i \in L, s \in S} J_{is}^L \hat{c}_s^\dagger(t)\hat{c}_i(t) e^{(i/2)\lambda_p + (i/2)\lambda_e \epsilon_i} + J_{is}^L \hat{c}_s(t)\hat{c}_i^\dagger(t) e^{(-i/2)\lambda_p + (-i/2)\lambda_e \epsilon_i}$$

$$+ \sum_{i \in R, s \in S} J_{is}^R \hat{c}_s^\dagger(t)\hat{c}_i(t) + J_{is}^R \hat{c}_s(t)\hat{c}_i^\dagger(t). \tag{C14}$$

Substituting Eq. (C14) into the quantum master equation (Eq. (C13)), the quantum master equation can be expressed by four nonzero terms, labeled by $K$.

$$\dot{\hat{\rho}}_S(\lambda, t) = -i[\hat{H}_S, \hat{\rho}_S(\lambda, t)] + \sum_{ss'} \int_0^t d\tau (T_1 + T_2 + T_3 + T_4). \tag{C15}$$

$$T_1 = -\text{Tr}_{Res}\{\hat{V}_\lambda^s \hat{V}_\lambda^{s'}(-\tau)\hat{\rho}_{Res}^{eq}\hat{\rho}_S(\lambda, t)\}$$

$$= -\text{Tr}_{Res}\left\{\sum_i J_{is}^L J_{is'}^L \hat{c}_i \hat{c}_i^\dagger(-\tau) \hat{\rho}_{Res}^{eq}\right\} \hat{c}_s^\dagger \hat{c}_{s'}(-\tau)\hat{\rho}_S(\lambda, t)$$

$$- \text{Tr}_{Res}\left\{\sum_i J_{is}^L J_{is'}^L \hat{c}_i^\dagger \hat{c}_i(-\tau) \hat{\rho}_{Res}^{eq}\right\} \hat{c}_s \hat{c}_{s'}^\dagger(-\tau)\hat{\rho}_S(\lambda, t)$$

$$- \text{Tr}_{Res}\left\{\sum_i J_{is}^R J_{is'}^R \hat{c}_i \hat{c}_i^\dagger(-\tau) \hat{\rho}_{Res}^{eq}\right\} \hat{c}_s^\dagger \hat{c}_{s'}(-\tau)\hat{\rho}_S(\lambda, t)$$

$$- \text{Tr}_{Res}\left\{\sum_i J_{is}^R J_{is'}^R \hat{c}_i^\dagger \hat{c}_i(-\tau) \hat{\rho}_{Res}^{eq}\right\} \hat{c}_s \hat{c}_{s'}^\dagger(-\tau)\hat{\rho}_S(\lambda, t)$$

$$T_2 = -\text{Tr}_{Res}\{\hat{\rho}_{Res}^{eq}\hat{\rho}_S(\lambda, t)\hat{V}_{-\lambda}^s(-\tau)\hat{V}_{-\lambda}^{s'}\}$$

$$= \text{Tr}_{Res}\left\{\hat{\rho}_{Res}^{eq} \sum_i J_{is}^L J_{is'}^L \hat{c}_i(-\tau)\hat{c}_i^\dagger\right\} \hat{\rho}_S(\lambda, t)\hat{c}_s^\dagger(-\tau)\hat{c}_{s'}$$

$$+ \text{Tr}_{Res}\left\{\hat{\rho}_{Res}^{eq} \sum_i J_{is}^L J_{is'}^L \hat{c}_i^\dagger(-\tau)\hat{c}_i\right\} \hat{\rho}_S(\lambda, t)\hat{c}_s(-\tau)\hat{c}_{s'}^\dagger$$

$$+ \text{Tr}_{Res}\left\{\hat{\rho}_{Res}^{eq} \sum_i J_{is}^R J_{is'}^R \hat{c}_i(-\tau)\hat{c}_i^\dagger\right\} \hat{\rho}_S(\lambda, t)\hat{c}_s^\dagger(-\tau)\hat{c}_{s'}$$

$$+ \text{Tr}_{Res}\left\{\hat{\rho}_{Res}^{eq} \sum_i J_{is}^R J_{is'}^R \hat{c}_i^\dagger(-\tau)\hat{c}_i\right\} \hat{\rho}_S(\lambda, t)\hat{c}_s(-\tau)\hat{c}_{s'}^\dagger$$

$$T_3 = \text{Tr}_{Res}\{\hat{V}_\lambda^s \hat{\rho}_{Res}^{eq} \hat{\rho}_S(\lambda,t) \hat{V}_{-\lambda}^{s'}(-\tau)\}$$

$$= \text{Tr}_{Res}\left\{\sum_i J_{is}^L J_{is'}^L \hat{c}_i \hat{\rho}_{Res}^{eq} \hat{c}_i^\dagger(-\tau) e^{i\lambda_p + i\lambda_e \epsilon_i}\right\} \hat{c}_s^\dagger \hat{\rho}_S(\lambda,t) \hat{c}_{s'}(-\tau)$$

$$+ \text{Tr}_{Res}\left\{\sum_i J_{is}^L J_{is'}^L \hat{c}_i^\dagger \hat{\rho}_{Res}^{eq} \hat{c}_i(-\tau) e^{-i\lambda_p - i\lambda_e \epsilon_i}\right\} \hat{c}_s \hat{\rho}_S(\lambda,t) \hat{c}_{s'}^\dagger(-\tau)$$

$$+ \text{Tr}_{Res}\left\{\sum_i J_{is}^R J_{is'}^R \hat{c}_i \hat{\rho}_{Res}^{eq} \hat{c}_i^\dagger(-\tau)\right\} \hat{c}_s^\dagger \hat{\rho}_S(\lambda,t) \hat{c}_{s'}(-\tau)$$

$$+ \text{Tr}_{Res}\left\{\sum_i J_{is}^R J_{is'}^R \hat{c}_i^\dagger \hat{\rho}_{Res}^{eq} \hat{c}_i(-\tau)\right\} \hat{c}_s \hat{\rho}_S(\lambda,t) \hat{c}_{s'}^\dagger(-\tau)$$

$$T_4 = \text{Tr}_{Res}\{\hat{V}_\lambda^s(-\tau) \hat{\rho}_{Res}^{eq} \hat{\rho}_S(\lambda,t) \hat{V}_{-\lambda}^{s'}\}$$

$$= \text{Tr}_{Res}\left\{\sum_i J_{is}^L J_{is'}^L \hat{c}_i(-\tau) \hat{\rho}_{Res}^{eq} \hat{c}_i^\dagger e^{i\lambda_p + i\lambda_e \epsilon_i}\right\} \hat{c}_s^\dagger(-\tau) \hat{\rho}_S(\lambda,t) \hat{c}_{s'}$$

$$+ \text{Tr}_{Res}\left\{\sum_i J_{is}^L J_{is'}^L \hat{c}_i^\dagger(-\tau) \hat{\rho}_{Res}^{eq} \hat{c}_i e^{-i\lambda_p - i\lambda_e \epsilon_i}\right\} \hat{c}_s(-\tau) \hat{\rho}_S(\lambda,t) \hat{c}_{s'}^\dagger$$

$$+ \text{Tr}_{Res}\left\{\sum_i J_{is}^R J_{is'}^R \hat{c}_i(-\tau) \hat{\rho}_{Res}^{eq} \hat{c}_i^\dagger\right\} \hat{c}_s^\dagger(-\tau) \hat{\rho}_S(\lambda,t) \hat{c}_{s'}$$

$$+ \text{Tr}_{Res}\left\{\sum_i J_{is}^R J_{is'}^R \hat{c}_i^\dagger(-\tau) \hat{\rho}_{Res}^{eq} \hat{c}_i\right\} \hat{c}_s(-\tau) \hat{\rho}_S(\lambda,t) \hat{c}_{s'}^\dagger$$

Set the equilibrium functions for reservoirs $L$ and $R$:

$$\alpha_{ss'}^X(\tau) = \sum_{i \in X = L,R} J_{is}^X (J_{is'}^X) \text{Tr}\{\hat{c}_i(\tau) \hat{c}_i^\dagger \hat{\rho}_{Res}^{eq}\} = \sum_{i \in X = L,R} J_{is}^X (J_{is'}^X) \text{Tr}\{\hat{c}_i \hat{c}_i^\dagger(-\tau) \hat{\rho}_{Res}^{eq}\}$$

$$\beta_{ss'}^X(\tau) = \sum_{i \in X = L,R} J_{is}^X (J_{is'}^X) \text{Tr}\{\hat{c}_i^\dagger(\tau) \hat{c}_i \hat{\rho}_{Res}^{eq}\} = \sum_{i \in X = L,R} J_{is}^X (J_{is'}^X) \text{Tr}\{\hat{c}_i^\dagger \hat{c}_i(-\tau) \hat{\rho}_{Res}^{eq}\}$$

$$\alpha_{ss'}(\tau) = \alpha_{ss'}^L(\tau) + \alpha_{ss'}^R(\tau) \qquad \beta_{ss'}(\tau) = \beta_{ss'}^L(\tau) + \beta_{ss'}^R(\tau) \qquad (C16)$$

Using the rule of trace calculation, $\text{Tr}(ABC) = \text{Tr}(CAB) = \text{Tr}(BCA)$, and the above functions, we can get a simple form of Eq. (C15).

$$\dot{\hat{\rho}}_S(\lambda,t) = -i[\hat{H}_S, \hat{\rho}_S(\lambda,t)]$$

$$+\sum_{ss'}\int_0^t d\tau[-\alpha_{ss'}(\tau)\hat{c}_s^\dagger \hat{c}_{s'}(-\tau)\hat{\rho}_S(\lambda,t) - \beta_{ss'}(\tau)\hat{c}_s \hat{c}_{s'}^\dagger(-\tau)\hat{\rho}_S(\lambda,t) - \alpha_{ss'}(-\tau)\hat{\rho}_S(\lambda,t)\hat{c}_s^\dagger(-\tau)\hat{c}_{s'}$$

$$- \beta_{ss'}(-\tau)\hat{\rho}_S(\lambda,t)\hat{c}_s(-\tau)\hat{c}_{s'}^\dagger$$

$$+ (\beta_{ss'}^L(-\tau)e^{i\lambda_p + i\lambda_e \epsilon_i} + \beta_{ss'}^L(-\tau))\hat{c}_s^\dagger \hat{\rho}_S(\lambda,t)\hat{c}_{s'}(-\tau)$$

$$+ (\alpha_{ss'}^L(-\tau)e^{-i\lambda_p - i\lambda_e \epsilon_i} + \alpha_{ss'}^R(-\tau))\hat{c}_s \hat{\rho}_S(\lambda,t)\hat{c}_{s'}^\dagger(-\tau)$$

$$+ (\beta_{ss'}^L(\tau)e^{i\lambda_p + i\lambda_e \epsilon_i} + \beta_{ss'}^R(\tau))\hat{c}_s^\dagger(-\tau)\hat{\rho}_S(\lambda,t)\hat{c}_{s'}$$

$$+ (\alpha_{ss'}^L(\tau)e^{-i\lambda_p - i\lambda_e \epsilon_i}$$

$$+ \alpha_{ss'}^R(\tau))\hat{c}_s(-\tau)\hat{\rho}_S(\lambda,t)\hat{c}_{s'}^\dagger]. \tag{C17}$$

Next, we apply the Markovian approximation (the upper time limit in Eq. (C17) becomes infinity) and the rotation wave approximation (ignoring the oscillation under the long-time average, which equals to make the reservoirs' equilibrium functions are diagonal in $s$). Then Eq. (C17) becomes

$$\dot{\hat{\rho}}_S(\lambda,t) = -i[\hat{H}_S, \hat{\rho}_S(\lambda,t)]$$

$$+\sum_s \int_0^\infty d\tau[-\alpha_s(\tau)\hat{c}_s^\dagger \hat{c}_s(-\tau)\hat{\rho}_S(\lambda,t) - \beta_s(\tau)\hat{c}_s \hat{c}_s^\dagger(-\tau)\hat{\rho}_S(\lambda,t) - \alpha_s(-\tau)\hat{\rho}_S(\lambda,t)\hat{c}_s^\dagger(-\tau)\hat{c}_s$$

$$- \beta_s(-\tau)\hat{\rho}_S(\lambda,t)\hat{c}_s(-\tau)\hat{c}_s^\dagger + \left(\beta_s^L(-\tau)e^{i\lambda_p + i\lambda_e \epsilon_i} + \beta_s^R(-\tau)\right)\hat{c}_s^\dagger \hat{\rho}_S(\lambda,t)\hat{c}_s(-\tau)$$

$$+ \left(\alpha_s^L(-\tau)e^{-i\lambda_p - i\lambda_e \epsilon_i} + \alpha_s^R(-\tau)\right)\hat{c}_s \hat{\rho}_S(\lambda,t)\hat{c}_s^\dagger(-\tau)$$

$$+ \left(\beta_s^L(\tau)e^{i\lambda_p + i\lambda_e \epsilon_i} + \beta_s^R(\tau)\right)\hat{c}_s^\dagger(-\tau)\hat{\rho}_S(\lambda,t)\hat{c}_s$$

$$+ \left(\alpha_s^L(\tau)e^{-i\lambda_p - i\lambda_e \epsilon_i}\right.$$

$$\left.+ \alpha_s^R(\tau)\right)\hat{c}_s(-\tau)\hat{\rho}_S(\lambda,t)\hat{c}_s^\dagger]. \tag{C18}$$

Additionally, using the relation, $\hat{c}_s^\dagger = e^{i\hat{H}\tau}\hat{c}_s^\dagger e^{-i\hat{H}\tau} = \hat{c}_s^\dagger + [i\hat{H}\tau, \hat{c}_s^\dagger] + \frac{1}{2!}[i\hat{H}\tau, [i\hat{H}\tau, \hat{c}_s^\dagger]] \cdots$, one can easily find $\hat{c}_s^\dagger(\tau) = e^{i\epsilon_s \tau}\hat{c}_s^\dagger$ and $\hat{c}_s(\tau) = e^{-i\epsilon_s \tau}\hat{c}_s$. Substitute them into Eq. (C18) and move the exponential factors into the trace term for reservoirs. After trace process, by using the infinite integral of the complex exponential, $\int_0^\infty e^{i(\epsilon_i - \epsilon_s)} = \pi\delta(\epsilon_i - \epsilon_s)$, we can obtain a boson distribution function related to the system's energy level $\epsilon_s$. Finally, replacing the delta function with an assumed density of states $g$, Eq. (C18) becomes

$$\dot{\hat{\rho}}_S(\lambda,t) = -i[\hat{H}_S, \hat{\rho}_S(\lambda,t)]$$

$$= 2\pi g \sum_s -[|J_s^L|^2(n_L(\epsilon_s) + 1) + |J_s^R|^2(n_R(\epsilon_s) + 1)]\hat{c}_s^\dagger \hat{c}_s \hat{\rho}_S(\lambda,t)$$

$$- [|J_s^L|^2 n_L(\epsilon_s) + |J_s^R|^2 n_R(\epsilon_s)]\hat{c}_s \hat{c}_s^\dagger \hat{\rho}_S(\lambda,t)$$

$$+ [|J_s^L|^2 n_L(\epsilon_s)e^{i\lambda_p + i\lambda_e \epsilon_i} + |J_s^R|^2 n_R(\epsilon_s)]\hat{c}_s^\dagger \hat{\rho}_S(\lambda,t)\hat{c}_s$$

$$+ [|J_s^L|^2(n_L(\epsilon_s) + 1)e^{-i\lambda_p - i\lambda_e \epsilon_i}$$

$$+ |J_s^R|^2(n_R(\epsilon_s) + 1)]\hat{c}_s \hat{\rho}_S(\lambda,t)\hat{c}_s^\dagger, \tag{C19}$$

where $n(\epsilon_s)$ is the distribution function. $n(\epsilon_s)$ and $n(\epsilon_s) + 1$ in Eq. (B19) are the manifestations of the particle number operator $\hat{c}^\dagger \hat{c}$ and its commuted operator $\hat{c}\hat{c}^\dagger$.

## Appendix D: Asymptotic results of the infinite matrix $M$

Analytically solving the tridiagonal infinite matrix (Eq. (29)) poses a significant challenge. Fortunately, with the increasing size of matrix $M$, its cumulants in a unit time gradually converge toward a limit value, expressed by $\gamma$ and $n$. Here, we give the asymptotic behaviors concerning the photon currents, employing the parameters in the two-reservoir system: $T_L = -6$, $\mu_L = 5$, $T_R = -3$, $\mu_R = 6$, $t' = 0.05$. The pertinent results are displayed in Appendix Fig. 1. Both the first and second cumulants progressively approach the limit expression as matrix $M$ grows larger. Notably, these limit expressions are the analogs of that for electron transport, differing in the mapping of $\hat{c}\hat{c}^\dagger$ ($n-1$ for electrons, and $n+1$ for photons) due to the commutative and anti-commutative relations. Thus, the limit expressions for photon transport are

$$\langle I \rangle = \sum_s \frac{\gamma_s^L \gamma_s^R}{\gamma_s^L + \gamma_s^R} \times [n_L(\epsilon_s) - n_R(\epsilon_s)] \tag{D1}$$

$$\Delta I = \sum_s \frac{\gamma_s^L \gamma_s^R}{\gamma_s^L + \gamma_s^R} \left( \frac{\gamma_s^L \gamma_s^R}{\gamma_s^L + \gamma_s^R} - 1 \right) [n_L(\epsilon_s) - n_R(\epsilon_s)]^2 + \frac{\gamma_s^L \gamma_s^R}{\gamma_s^L + \gamma_s^R} [n_L(\epsilon_s)(n_L(\epsilon_s) + 1)$$

$$+ n_R(\epsilon_s)(n_R(\epsilon_s) + 1)] \#(C2)$$

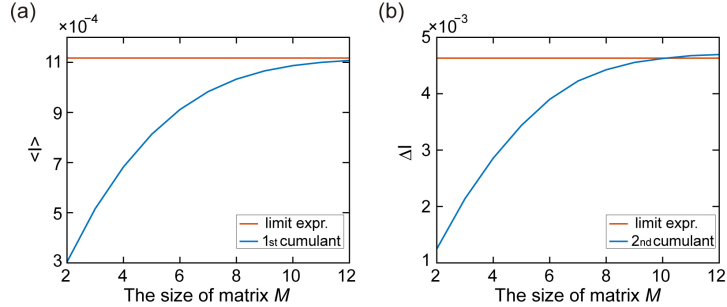

FIG. 5. Asymptotic behaviors of the cumulants $(-i)^n \frac{\partial^n}{\partial \lambda^n} v_{max}\big|_{\lambda=0}$. (a) gives the dependence of the first cumulant on the size of the matrix $M$, while (b) indicates the second cumulant versus the size of the matrix $M$.

## Appendix References